\documentclass[journal]{article}
\usepackage{color, anysize}
\usepackage{algorithmic}
\usepackage{algorithm}
\usepackage{graphicx}
\marginsize{0.7in}{0.5in}{0.5in}{1in}

\begin{document}

\title{Synthesis of the Optimal 4-bit Reversible Circuits}

\author{Oleg~Golubitsky\thanks{O. Golubitsky is with the Google Inc., Waterloo, ON, Canada.},
		Sean~M.~Falconer\thanks{S. M. Falconer is with the School of Medicine, Stanford University, 
Stanford, CA, USA.},
		Dmitri~Maslov\thanks{D. Maslov is with the Institute for Quantum Computing, University of Waterloo,
Waterloo, ON, N2L 3G1, Canada, email: {dmitri.maslov@gmail.com}.}
}

\maketitle

\begin{abstract}


Optimal synthesis of reversible functions is a non-trivial problem.  One of the major limiting
factors in computing such circuits is the sheer number of reversible functions.  Even restricting
synthesis to 4-bit reversible functions results in a huge search space (16! $\approx$ $2^{44}$
functions).  The output of such a search alone, counting only the space required to list Toffoli
gates for every function, would require over 100 terabytes of storage.

In this paper, we present an algorithm, that synthesizes an optimal
circuit for any 4-bit reversible specification. We employ several techniques to make the problem
tractable. We report results from several experiments, including synthesis of random 4-bit permutations,
optimal synthesis of all 4-bit linear reversible circuits,
synthesis of existing benchmark functions, and distribution of optimal circuits.  Our results
have important implications for the design and optimization of quantum circuits, testing circuit
synthesis heuristics, and performing experiments in the area of quantum information processing.

\end{abstract}

\section{Introduction}
 
To the best of our knowledge, at present, physically reversible technologies are found only in 
the quantum domain \cite{bk:nc}. However, ``quantum'' unites several technological approaches 
to information processing, including ion traps, optics, superconducting, spin-based and 
cavity-based technologies \cite{bk:nc}. Of those, trapped ions \cite{ar:s-k} and
liquid state NMR (Nuclear Magnetic Resonance) \cite{ar:n} are two of the most
developed quantum technologies targeted for computation in the circuit model (as opposed to communication
or adiabatic computing). These technologies allow computations over a set of 8 qubits and 
12 qubits, correspondingly.

Reversible circuits are an important class of computations that need to be performed efficiently for the purpose of efficient 
quantum computation. Multiple quantum algorithms contain arithmetic units such as adders, multiplies, exponentiation, 
comparators, quantum register shifts and permutations, that are best viewed as reversible circuits.  Moreover, reversible 
circuits are indispensable in quantum error correction \cite{bk:nc}. Often, the efficiency of the reversible implementation is 
the bottleneck of a quantum algorithm (e.g., integer factoring and discrete logarithm \cite{ar:s}) or even a class of quantum 
circuits (e.g., stabilizer circuits \cite{ar:ag}).

In this paper, we present an algorithm that finds optimal circuit implementations for 4-bit reversible functions. The algorithm 
has a number of potential uses and implications.

One major implication of this work is that it will help physicists with experimental design, since fore-knowledge of the optimal 
circuit implementation aids in the control over quantum mechanical systems.
The control of quantum mechanical systems is very difficult, and as a result experimentalists are always looking for the best 
possible implementation. Having an optimal implementation helps to improve experiments or show that more control over a physical 
system needs to be established before a certain experiment could be performed.

A second important contribution is due to the
efficiency of our implementation---$0.01$ seconds per synthesis of an optimal 4-bit reversible circuit.  The algorithm could 
easily be integrated as part of peephole optimization, such as the one presented in \cite{ar:pspmh}.

Furthermore, our implementation allows us to propose a subset of optimal implementations that may be used to test heuristic 
synthesis algorithms. Currently, similar tests are performed by comparison to optimal 3-bit implementations. The best heuristic 
solutions have very tiny overhead, making such a test hard to improve. As such, it would help to replace this test with a more 
difficult one that allows more room for improvement.

Finally, due to the effectiveness of our approach, we are able to report new optimal implementations for small benchmark 
functions, approximate $L(4)$, the number of reversible gates required to implement a reversible 4-bit function, approximate the 
average number of gates required to implement a 4-bit permutation, and show the distribution of the number of permutations that 
may be implemented with $0..9$ gates.

\section{Preliminaries}

\begin{figure}
\centering
\includegraphics[height=22mm]{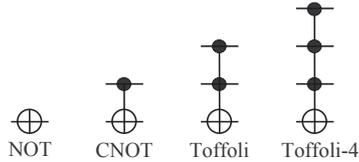}
\caption{NOT, CNOT, Toffoli, and Toffoli-4 gates.}
\label{fig1}
\end{figure}

In this paper, we consider circuits with NOT, CNOT, Toffoli (TOF), and Toffoli-4 (TOF4) gates defined as follows:
\begin{itemize}
\item NOT$(a):\; a\mapsto a\oplus 1$;
\item CNOT$(a,b):\; a,b\mapsto a, b\oplus a$;
\item TOF$(a,b,c):\; a,b,c\mapsto a, b, c\oplus ab$;
\item TOF4$(a,b,c,d):\; a,b,c,d\mapsto a, b, c, d\oplus abc$;
\end{itemize}
\noindent where $\oplus$ denotes an EXOR operation and concatenation is Boolean AND; see Figure \ref{fig1} for illustration. 
These gates are used widely in quantum circuit construction, and have been demonstrated experimentally in multiple quantum 
information processing
proposals \cite{bk:nc}. In particular, CNOT is a very popular gate among experimentalists, frequently used to demonstrate 
control over a multiple-qubit quantum mechanical system. Since quantum circuits describe time evolution of a quantum mechanical 
system where individual ``wires'' represent physical instances, and time propagates from left to right, this imposes 
restrictions on the circuit topology. In particular, quantum and reversible circuits are strings of gates. As a result, 
feed-back (time wrap) is not allowed and there may be no fan-out (mass/energy conservation).

In this paper, we are concerned with searching for circuits requiring a minimal number of gates. Our focus is the proof of 
principle, i.e., showing that any optimal 4-bit reversible function may be synthesized efficiently, rather than attempting to 
report optimal implementations for a number of potentially plausible cost metrics. In fact, our implementation allows other 
circuit cost metrics to be considered, as discussed in Section \ref{s:cfr}.

In related work, there have been a few attempts to synthesize optimal reversible circuits with more than three inputs. 
Gro{\ss}e {\it et al.} \cite{co:gcdd} employ SAT-based technique to synthesize provably optimal circuits for some small 
parameters.
However, their implementation quickly runs out of resources. The longest optimal circuit they report contains 11 gates. The 
latter took 21,897.3 seconds to synthesize---same function that the implementation we report in this paper synthesized in 
.000106 seconds, see Table \ref{tab3}.
Prasad {\it et al.} \cite{ar:pspmh} used breadth first search to synthesize 26,000,000 optimal 4-bit reversible circuits with up 
to 6 gates in 152 seconds. We extend this search into finding 117,798,040,190 optimal circuits with up to 9 gates in 10,549 
seconds. This is over 65 times faster and 4,500 times more than reported in \cite{ar:pspmh}.
Yang {\it et al.} \cite{ar:yshp} considered short optimal reversible 4-bit circuits composed with NOT, CNOT, and Peres 
\cite{ar:p} gates. They were able to synthesize optimal circuits for even permutations requiring no more than 12 gates. This 
amounts to approximately one quarter of the number of all 4-bit reversible functions. Our implementation allows optimal 
synthesis of any 4-bit reversible function, and it is much faster. 

\subsection{Motivating Example}
\begin{figure}
\centering
\includegraphics[height=22mm]{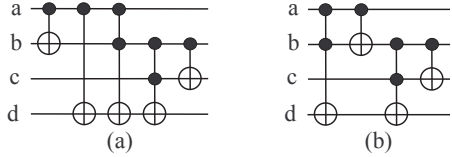}
\caption{(a) a suboptimal and (b) an optimal circuit for 1-bit full adder.}
\label{fig}
\end{figure}

Consider the two reversible circuit implementations in Figure~\ref{fig} of a 1-bit full adder. This elementary function/circuit 
serves as a building block for constructing integer adders. The famous Shor's integer factoring algorithm is dominated by adders 
like this.  As such, the complexity of an elementary 1-bit adder circuit largely affects the efficiency of factoring an integer 
number with a quantum algorithm. It is thus important to have a well-optimized 
implementation of a 1-bit adder, as well as other similar small quantum circuit building blocks. 

In this paper, we consider the synthesis of optimal circuits, i.e., we provably find the best possible implementation. Using 
optimal implementations of circuits potentially increases the efficiency of quantum algorithms and helps to reduce the 
difficulty with controlling quantum experiments.

\section{Algorithm and Implementation}

We first outline our algorithm and then discuss it in detail in the follow up subsections.

There are $N = 2^n!$ reversible $n$-variable functions. The most obvious
approach to the synthesis of all optimal implementations is to compute all
optimal circuits and store them for later look-up. However, this is extremely inefficient.  This is because such an
approach requires $\Omega(N)$ space and, as a result, at least $\Omega(N)$ time. These space and
time estimates are lower bounds, because, for instance, storing an optimal circuit requires more
than a constant number of bits, but for simplicity, let us assume these figures are exact.
Despite considering both figures for space and time unpractical,
we use this simple idea as our starting point.

We first improve the space
requirement by observing that if one synthesized all halves of all optimal circuits,
then it is possible to search through this set to find both halves of any optimal
circuit. It can be shown that the space requirement for storing halves has a lower bound of
$\Omega(\sqrt{N})$. However, searching for two halves potentially requires a runtime
on the order of the square of the search space, $\Omega\left((\sqrt{N})^2\right)=\Omega(N)$,
a figure for runtime that we deemed inefficient. Our second improvement is
thus to use a hash table to store the optimal halves. This reduces the runtime
to soft $\Omega(\sqrt{N})$. While this lower bound does not necessarily imply
that the actual complexity is lower than $O(N)$, this turns out to be the case,
because the set of optimal halves is indeed much smaller than the set of all 
optimal circuits (an analytic estimate for the relative size of the former set 
is hard to obtain, though). Cumulatively, these two improvements reduce $\Omega(N)$ 
space and $\Omega(N)$ time requirement to $O(\#{\rm halves}(N))$ space and
soft $O(\#{\rm halves}(N))$ time requirement. These reductions almost suffice
to make the search possible using modern computers.

Our last step, apart from careful coding, that made the search possible is the
reduction of the space requirement (with consequent improvement for runtime) by a
constant of almost 48 via exploiting the following two features. First, simultaneous
input/output relabeling, of which there are at most 24 (=4!) different ones, does not change the
optimality of a circuit. And second, if an optimal circuit is found for a function $f$, an optimal circuit for the inverse 
function, $f^{-1}$, 
can be obtained by reversing the optimal circuit for $f$. This allows to additionally ``pack'' up
to twice as many functions into one circuit. The cumulative improvement
resulting from these two observations, is by a factor of almost $2\times 24=48$.
Due to symmetries, the actual number is slightly less. See Table \ref{tab1}
(column 2 versus column 3) for exact comparison.

\subsection{The search-and-lookup algorithm}

For brevity, let the
size of a reversible function mean the minimal number of gates required
to implement it. Using breadth-first search, we can generate
the smallest circuits for all reversible functions of size at most $k$,
for a certain value of $k$.
(This can be done in advance, on a larger machine, and need not be repeated
for each reversible function.)

Assume that the given function $f$, for which we need to synthesize
a minimal circuit, has size at most $2k$. We can first check whether
$f$ is among the known functions of size at most $k$ and, if so,
output the corresponding minimal circuit. If not, then the size of
$f$ is between $k+1$ and $2k$, inclusive, and there exist reversible
functions $g$ and $h$ of size $k$ and at most $k$, respectively, such
that $f=h\circ g$. If we find such $g$ of the smallest size, then we
can obtain the smallest circuit for $f$ by composing the circuits for
$g$ and $h$.

Multiplying the above equality by $g^{-1}$, we obtain $f\circ g^{-1}=h$.
Observe that $g^{-1}$ has the same size as $g$. Therefore, by trying
all functions $g$ of size $1,2,\ldots,k$ until we find one such
that $f\circ g$ has size $k$, we can find a $g$ of the smallest size.

The above algorithm involves sequential access to the functions of size
at most $k$ and their minimal circuits and a membership test among
functions of size $k$. Since the latter test must be fast and requires
random memory access, we need to store all functions of size $k$ in memory.
Thus, the amount of available RAM imposes an upper bound on $k$.

In practice, we store a 4-bit reversible function using a 64-bit word,
because this allows for an efficient implementation of functional
composition, inversion, and other necessary operations.
On a typical PC with 4GB of RAM, we can store all functions for $k=6$.
This means that we can apply the above search algorithm only to functions
of size at most 12. Unfortunately, this will not cover all 4-bit reversible
functions. Therefore, further reduction of memory usage is necessary.

\subsection{Symmetries}

A significant reduction of the search space can be achieved by taking into
account the following symmetries of circuits:

\begin{enumerate}
  \item Simultaneous relabeling of inputs and outputs. Given an optimal circuit
implementing a 4-bit reversible function $f$ with inputs $x_0,x_1,x_2,x_3$
and outputs $y_0,y_1,y_2,y_3$ and a permutation $\sigma:\{0,1,2,3\}\to\{0,1,2,3\},$
we can construct a new circuit by relabeling the inputs and outputs into
$x_{\sigma(0)},x_{\sigma(1)},x_{\sigma(2)},x_{\sigma(3)}\;{\rm and}\;
y_{\sigma(0)},y_{\sigma(1)},y_{\sigma(2)},y_{\sigma(3)},$ respectively.
Then the new circuit will provide a minimal implementation of the corresponding
reversible function $f_\sigma$. Indeed, if it is not minimal and
there is an implementation of $f_\sigma$ by a circuit with a smaller number of gates,
we can relabel the inputs and outputs of this implementation with $\sigma^{-1}$
and obtain a smaller circuit implementing the original function $f$. This contradicts
the assumption that the original circuit for $f$ is optimal.

Given $f$ and $\sigma$, a formula for $f_\sigma$ can be easily obtained.
Observe that the mapping
$x_0,x_1,x_2,x_3 \mapsto x_{\sigma(0)},x_{\sigma(1)},x_{\sigma(2)},$ $x_{\sigma(3)}$
is a 4-bit reversible function, which we denote by $g_\sigma$. The mapping
$y_{\sigma(0)},y_{\sigma(1)},y_{\sigma(2)},y_{\sigma(3)} \mapsto y_0,y_1,y_2,y_3$
is then given by the inverse, $g_\sigma^{-1}$. Therefore, the four bit values
$y_0,y_1,y_2,y_3$ of $f_\sigma$ on a four-bit tuple $x_0,x_1,x_2,x_3$ can be
 obtained by applying first $g_\sigma$, then $f$, and finally $g_\sigma^{-1}$.
We obtain
$f_\sigma = g_\sigma^{-1} \circ f \circ g_\sigma.$
We call the set of functions $f_\sigma$ the {\it conjugacy class} of $f$
modulo simultaneous input/output relabelings.

Since there exist 24 permutations of 4 numbers, by choosing different permutations
$\sigma$, we obtain 24 functions of the above form $f_\sigma$ for a fixed function $f$.
Some of these functions may be equal, whence the size of the conjugacy class of $f$
may be smaller than 24. For example, if $f$=NOT$(a)$, then there exist only $4$ distinct
functions of the form $f_\sigma$ (counting $f$ itself). Our experiments show, however,
that for the vast majority of functions, the conjugacy classes are of size 24.

  \item Inversion. As mentioned above, if we know a minimal implementation for $f$,
then we know one for its inverse as well.
\end{enumerate}

Note that conjugation and inversion commute:
$$(g_\sigma^{-1}\circ f \circ g_\sigma)^{-1} = g_\sigma^{-1}\circ f^{-1} \circ g_\sigma.$$
For a function $f$, consider the union of the two conjugacy classes of $f$ and $f^{-1}$.
Call the elements of this union {\it equivalent} to $f$. It follows that equivalent
functions have the same size. Moreover, since gates are idempotent (i.e., equal
to their own inverses) and their conjugacy classes consist of gates, if we know a
minimal circuit for $f$, we can easily obtain one for any function in the equivalence class of $f$.
Formally, if $f=g_1\circ\ldots\circ g_n$, where $n$ is the size of $f$ and
$g_i$ are gates, then $f^{-1} = g_n\circ\ldots\circ g_1$, and if
$f'=g_\sigma^{-1}\circ f\circ g_\sigma$, then
$f'=g_1'\circ\ldots\circ g_n'$, where $g_i' = g_\sigma^{-1}\circ g_i\circ g_\sigma$
are also gates.
Our experiments show that a vast majority of functions have 48 distinct equivalent
functions. This fact can reduce the search space by almost a factor of 48 as follows.

For a function $f$, define the canonical representative of its equivalence class.
A convenient canonical representative can be obtained by introducing the lexicographic
order on the set of 4-bit reversible functions, considered as permutations of
$\{0,1,2,\ldots,15\}$ and encoded accordingly by the sequence $f(0),f(1),\ldots,f(15)$,
and choosing the function whose corresponding sequence is lexicographically smallest.
Now, instead of storing all functions of size at most $k$, store the canonical
representative for each equivalence class. This will reduce the storage size
by almost a factor of 48.
Then, we use the Algorithm~\ref{slsym} to search for a minimal circuit for a
given reversible function $f$.

\begin{algorithm}[t]
\caption{Minimal circuit.}
\label{slsym}
\begin{algorithmic}
\REQUIRE   Reversible function $f$ of size at most $L$.\\
 Hash table $H$ containing canonical representatives of all equivalence
classes of functions of size at most $k$ and the last gates of their minimal
circuits, $k\ge L/2$.\\
 Lists $A_i$, $1\le i\le L-k$, of all functions of size $i$.
\ENSURE A minimal circuit $c$ for $f$.

\IF{$f=\;$IDENTITY}
  \RETURN{empty circuit}
\ENDIF

\STATE $E_f\leftarrow $ equivalence class of $f$
\STATE $\bar f\leftarrow$ canonical representative of $E_f$
\IF{$\bar f\in H$}
  \STATE $\bar\lambda\leftarrow$ last gate of $\bar f$
  \IF{$f$ is a conjugate of $\bar f$}
    \STATE let $f = g_\sigma^{-1}\circ \bar f \circ g_\sigma$
    \STATE $\lambda\leftarrow g_\sigma^{-1}\circ\bar\lambda \circ g_\sigma$
    \STATE $c\leftarrow$ minimal circuit for $f\circ\lambda$
    \RETURN {$c\circ \lambda$}
  \ELSE
    \STATE let $f = g_\sigma^{-1}\circ \bar f^{-1} \circ g_\sigma$
    \STATE $\lambda\leftarrow g_\sigma^{-1}\circ\bar\lambda \circ g_\sigma$
    \STATE $c\leftarrow$ minimal circuit for $\lambda \circ f$
    \RETURN {$\lambda\circ c$}
  \ENDIF
\ENDIF

\FOR{$i=1$ to $L-k$}
  \FOR {$g\in A_i$}
    \STATE $h\leftarrow g\circ f$
    \STATE $E_h\leftarrow$ equivalence class of $h$
    \STATE $\bar h\leftarrow$ canonical representative of $E_h$
    \IF{$\bar h\in H$}
       \STATE $c_g\leftarrow$ minimal circuit for $g$
       \STATE $c_h\leftarrow$ minimal circuit for $h$
       \RETURN{$c_g^{-1}\circ c_h$}
    \ENDIF
  \ENDFOR
\ENDFOR

\RETURN{{\bf error:} size of $f$ is greater than $L$}

\end{algorithmic}

\end{algorithm}

The algorithm requires a hash table with canonical representatives of equivalence classes
of size at most $k$, together with the last gates of their minimal circuits, and lists
of all permutations of size at most $L-k$.
We have pre-computed the canonical representatives for $k=9$ using breadth-first search
(see Algorithm~\ref{bfs}).
For efficiency reasons, we store the last {\it or the first} gate of
a minimal circuit for each canonical representative. However, this information is clearly
sufficient to reconstruct the entire circuit and, in particular, the last gate.
Using this pre-computed data, the hash table and the lists of all permutations of size
at most $L-k$ are formed at the start-up. An implementation storing only the hash table is
possible. Such an implementation will require less RAM memory, but it will be slower. We
decided to focus on higher speed, because Table \ref{tab2} indicates that we do not
need to be able to search optimal circuits requiring up to 18 ($=9\times 2$) gates,
which we could do otherwise by storing only the hash table.

\begin{algorithm}[t]
\caption{Breadth-first search.}
\label{bfs}
\begin{algorithmic}
  \REQUIRE $k$
  \ENSURE Lists $A_i$ of canonical representatives of size $\le k$;\\
    Hash table $H$ with these canonical representatives and their first or last gates.\\
  \STATE Let $H$ be a hash table (keys are functions, values are gates)
  \STATE $H$.insert(IDENTITY, HAS\_NO\_GATES)
  \STATE $A_0\leftarrow\{{\rm IDENTITY}\}$
  \FOR {$i$ from $1$ to $k$}
    \FOR {$f\in A_{i-1} \cup \{a^{-1}\;|\;a\in A_{i-1}\}$}
      \FOR {all gates $\lambda$}
         \STATE $h\leftarrow f\circ\lambda$
         \STATE $E_h\leftarrow$ equivalence class of $h$
         \STATE $\bar h\leftarrow$ canonical representative of $E_h$
         \IF {$\bar h\not\in H$}
            \IF {$h$ is a conjugate of $\bar h$}
              \STATE let $h=g_\sigma^{-1}\circ \bar h\circ g_\sigma$
              \STATE $H$.insert($\bar h$, $g_\sigma^{-1}\circ \lambda\circ g_\sigma$, IS\_A\_LAST\_GATE)
            \ELSE
              \STATE let $h=g_\sigma^{-1}\circ \bar h^{-1}\circ g_\sigma$
              \STATE $H$.insert($\bar h$, $g_\sigma^{-1}\circ \lambda\circ g_\sigma$, IS\_A\_FIRST\_GATE)
            \ENDIF
            \STATE $A_i$.insert($\bar h$)
         \ENDIF
       \ENDFOR
     \ENDFOR
  \ENDFOR
\end{algorithmic}
\end{algorithm}

The correctness of Algorithm~\ref{slsym} is proved as follows. Suppose first that
the size of $f$ is at most $k$. The canonical representative $\bar f$
of its equivalence class will have the same size as $f$, so it will be
found in the hash table $H$. Since $\bar\lambda$ is the last gate of a
minimal circuit for $\bar f$, the size of $\bar f\circ\bar \lambda$ is
one less than the size of $\bar f$. The function $f\circ\lambda$
(computed if $f$ is a conjugate of $\bar f$) or
the function $\lambda\circ f$ (computed if $f$ is a conjugate of $\bar f^{-1}$)
is equivalent to $\bar f\circ\bar\lambda$ and therefore also is of size one
less than the size of $\bar f$. Therefore, the recursive call on that function
will terminate and return a minimal circuit, which we can compose with $\lambda$
(at the proper side) to obtain a minimal circuit for $f$. The depth of recursion
is equal to the size of $f$, and at each call we do one hash table lookup,
one computation of the canonical representative, and one conjugation of a gate
(the latter can be looked up in a small table). Thus, this part
of the algorithm requires negligible time.

If the size of $f$ is greater than $k$, but does not exceed $L$, then
$f=g_f\circ h$ for some $h$ of size $k$ and $g_f$ of size $i$, $1\le i\le L-k$.
Then $g=g_f^{-1}\in A_i$. Once the inner for-loop encounters this $g$, it
will return the minimal circuit for $f$, because both recursive calls are
for functions of size at most $k$. For a function $f$ of size $s>k$,
the number of iterations required to find the minimal circuit satisfies
$$\sum_{i=1}^{s-1-k}|A_i|<r\le \sum_{i=1}^{s-k}|A_i|.$$
At each iteration, one canonical representative is computed and looked up
in the hash table. Since the size of $A_i$ grows almost exponentially
(see Table~\ref{tab1}, left column), the search time will decrease
almost exponentially, and the storage will increase exponentially,
as $k$ increases. The timings for $k=8,9$ measured on two different
systems are summarized in Table~\ref{timings} (see Section~\ref{sec:performance} for machine details). The hash table loading
and overall memory usage times were 119 seconds, 3.5GB ($k=8$) and
1111 seconds, 43.04GB ($k=9$).

\begin{table}
\begin{center}
\caption{Average times of computing minimal circuits of sizes $0..14$ (in seconds).}
\label{timings}
\begin{tabular}{|l|rrr|}
\hline
Size \hfill $\setminus$ \hfill $k$ & 8 (CS2) & 8 (CS1) & 9 (CS1)\\
\hline
0 & $5.10\times 10^{-7}$ & $5.15\times 10^{-7}$ & $5.15\times 10^{-7}$ \\
1 & $8.70\times 10^{-7}$ & $8.80\times 10^{-7}$ & $8.80\times 10^{-7}$ \\
2 & $1.26\times 10^{-6}$ & $1.27\times 10^{-6}$ & $1.27\times 10^{-6}$ \\
3 & $1.66\times 10^{-6}$ & $1.69\times 10^{-6}$ & $1.68\times 10^{-6}$ \\
4 & $2.07\times 10^{-6}$ & $2.14\times 10^{-6}$ & $2.14\times 10^{-6}$ \\
5 & $2.47\times 10^{-6}$ & $2.52\times 10^{-6}$ & $2.52\times 10^{-6}$ \\
6 & $3.48\times 10^{-6}$ & $3.49\times 10^{-6}$ & $3.96\times 10^{-6}$ \\
7 & $4.22\times 10^{-6}$ & $4.11\times 10^{-6}$ & $4.85\times 10^{-6}$ \\
8 & $4.49\times 10^{-6}$ & $4.36\times 10^{-6}$ & $4.45\times 10^{-6}$ \\
9 & $1.07\times 10^{-5}$ & $1.18\times 10^{-5}$ & $5.65\times 10^{-6}$ \\
10& $2.28\times 10^{-4}$ & $2.77\times 10^{-4}$ & $1.79\times 10^{-5}$ \\
11& $4.27\times 10^{-3}$ & $5.20\times 10^{-3}$ & $2.38\times 10^{-4}$ \\
12& $6.30\times 10^{-2}$ & $7.66\times 10^{-2}$ & $3.74\times 10^{-3}$ \\
13& $4.91\times 10^{-1}$ & $5.98\times 10^{-1}$ & $3.18\times 10^{-2}$ \\
14& $4.38\times 10^0\;\,\,$ & $5.33\times 10^0\;\,\,$    & $3.26\times 10^{-1}$ \\
\hline
\end{tabular}
\end{center}
\end{table}

It follows from the above complexity analysis that the performance of the
following key operations affect the speed most:
\begin{itemize}
  \item composition of two functions ($f\circ g$) and inverse of a function ($f^{-1}$),
  \item computation of the canonical representative of an equivalence class,
  \item hash table lookup.
\end{itemize}
In the next Subsection we discuss an efficient implementation of these operations.

\subsection{Implementation details}

As mentioned above, a 4-bit reversible function can be stored in
a 64-bit word, by allocating 4 bits for each value of
$f(0),f(1),\ldots,f(15)$.
Then the composition of two functions can be computed in 94 machine
instructions using the algorithm {\tt composition} and
the inverse function can be computed in 59 machine instructions
using algorithm {\tt inverse}.

\begin{figure}[h]
\begin{verbatim}
unsigned64 composition(unsigned64 p, unsigned64 q) {
  unsigned64 d = (p & 15) << 2;
  unsigned64 r = (q >> p_i) & 15;
  p >>= 2; d = p & 60; r |= ((q >> d) & 15) << 4;
  p >>= 4; d = p & 60; r |= ((q >> d) & 15) << 8;
  p >>= 4; d = p & 60; r |= ((q >> d) & 15) << 16;
  ...
  p >>= 4; d = p & 60; r |= ((q >> d) & 15) << 60;
  return r;
}

unsigned64 inverse(unsigned64 p) {
  p >>= 2;
  unsigned64 q =  1 << (p & 60);
  p >>= 4;  q |=  2 << (p & 60);
  p >>= 4;  q |=  3 << (p & 60);
  ...
  p >>= 4;  q |= 15 << (p & 60);
  return q;
}

unsigned64 conjugate01(unsigned64 p) {
  p = (p & 0xF00FF00FF00FF00F)       |
     ((p & 0x00F000F000F000F0) << 4) |
     ((p & 0x0F000F000F000F00) >> 4);
  return (p & 0xCCCCCCCCCCCCCCCC)       |
        ((p & 0x1111111111111111) << 1) |
        ((p & 0x2222222222222222) >> 1);
}
\end{verbatim}
\end{figure}

In order to find the canonical representative in the equivalence class
of a function $f$, we compute $f^{-1}$, generate all conjugates
of $f$ and $f^{-1}$, and choose the smallest among the resulting 48
functions. Since every permutation of $\{0,1,2,3\}$ can be represented
as a product of transpositions $(0,1)$, $(1,2)$, and $(2,3)$,
the sequence of conjugates of $f$ by all 24 permutations can be
obtained through conjugating $f$ by these transpositions. These
conjugations can be performed in 14 machine instructions each
as in function {\tt conjugate01}.

Two functions can be compared lexicographically using a single
unsigned comparison of the corresponding two words. Thus, the canonical
representative can be computed using one inversion, $23\times 2 = 46$
conjugations by transpositions, and 47 comparisons, which totals
to 750 machine instructions.

For the fast membership test, we use a linear probing
hash table with Thomas Wang's hash function \cite{wang} (see algorithm {\tt hash64shift}).

\begin{figure}[h]
\begin{verbatim}
long hash64shift(long key)
{
  key = (~key) + (key << 21); // signed shift
  key = key ^ (key >>> 24);   // unsigned shift
  key = (key + (key << 3)) + (key << 8);
  key = key ^ (key >>> 14);
  key = (key + (key << 2)) + (key << 4);
  key = key ^ (key >>> 28);
  key = key + (key << 31);
  return key;
}
\end{verbatim}
\end{figure}

This function is well suited for our purposes: it is fast
to compute and distributes the permutations uniformly over the
hash table. The parameters of the hash tables storing
the canonical representatives of equivalence classes of size $k$,
for $k=7,8,9$ are shown in Table~\ref{hashtables}.

\begin{table}
\begin{center}
\caption{Parameters of linear hash tables storing canonical representatives.}
\label{hashtables}
\begin{tabular}{|l|ccc|}
\hline
\hfill $k$ & 7 & 8 & 9 \\
\hline 
Size$\vphantom{2^{25^{25}}}$ & $2^{25}$ & $2^{28}$ & $2^{32}$ \\
Memory Usage & 256 MB & 2 GB & 32 GB\\
Load Factor & 0.58 & 0.84 & 0.51 \\
Average Chain Length & 3.14 & 9.18 & 2.63\\
Maximal Chain Length & 92 & 754 & 86\\
\hline
\end{tabular}
\end{center}
\end{table}

\section{Performance and Results}
\label{sec:performance}
We performed several tests using two computer systems, $CS1$ and $CS2$.
$CS1$ is a high performance server with 16 AMD Opteron 2300 MHz processors,
64 GB RAM, and Seagate Barracuda ES2 SCSI 7200 RPM HDD running Linux.
$CS2$ is a laptop Sony VGN-NS190D with
Intel Core Duo 2000 GHz processor, 4 GB RAM, and a 5400 RPM SATA HDD running Linux.
The following subsections summarize the tests and results.

\subsection{Synthesis of Random Permutations} \label{ss:1}

\begin{table}
\centering
\begin{small}
\caption{Distribution of the number of gates required for 10,000,000 random 4-bit reversible functions.}
\label{tab2}
\begin{tabular}{|r|r|} \hline
Size  	& Functions 		\\  \hline
14		& 17,191 	\\
13		& 2,371,039 	\\
12		& 5,110,943		\\
11		& 2,051,507 		\\
10		& 392,108 		\\
9		& 50,861			\\
8		& 5,269 			\\
7		& 455 				\\
6		& 24				\\
5   	& 3         		\\ \hline
\end{tabular}
\end{small}
\end{table}

In this test, we generated 10,000,000 random uniformly distributed permutations using the Mersenne twister random number 
generator~\cite{ar:mn}. The test
was executed on $CS1$.
It took 104,616.716 seconds (about 29 hours) of user time and the
maximal RAM memory usage was 43.04GB. Note that 1111 seconds (approximately 18 minutes) were spent loading 
previously computed optimal circuits with up to 9 gates (see Subsection \ref{ss:2} for details) into RAM.
On average, it took only 0.01035 seconds to synthesize an optimal circuit for a permutation. The distribution 
of the circuit sizes is shown in Table \ref{tab2}.

Note, that the ratio of the number of random permutations requiring 9 gates to the number of all random 
permutations,  $\frac{50,861}{10M} \approx .005086$, is close to the ratio of the number of all permutations 
requiring 9 gates to the number of all permutations, $\frac{105,984,823,653}{16!} \approx .005066$. This implies 
that the weighted average over the random sample, equal to $11.94$ gates per circuit, must be close to the 
actual weighted average. We further use this random sample and the results of the optimal 3-bit circuit 
synthesis \cite{ar:spmh} to approximate the number of permutations requiring 10 through 17 gates, see Table \ref{tab1}.

We conjecture that there are no permutations requiring 17 gates, and unlikely many, if at all, that 
require 16 gates. This implies that our search may be performed on a machine capable of storing 
reduced optimal implementations with up to 8 gates, i.e., a machine with 4GB RAM. Further analysis 
suggests that the search for an optimal circuit will complete in the majority of cases (99.999$\%$ 
assuming uniform distribution) if one uses optimal circuits with at most 7 gates and stores only 
the hash table. Such a search requires slightly more than 256M of available RAM, and could be 
executed on an older machine.

\subsection{Distribution of Optimal Implementations} \label{ss:2}

\begin{table}[h]
\centering
\caption{Number of 4-bit permutations requiring 0..9 gates, and estimates of
the number of permutations requiring 10..17 gates.  Note, we were not able to estimate the number of functions for size 15 and 
16.}
\label{tab1}
\begin{tabular}{|r|r|r|} \hline
Size  	& Functions 		& Reduced \\
      	&           		& Functions \\ \hline
$\geq$17		& 0 & \\
16		& ??? & \\
15		& ??? & \\
14		& $\sim 3.60\times 10^{10}$ & \\
13		& $\sim 4.96\times 10^{12}$ & \\
12		& $\sim 1.07\times 10^{13}$ & \\
11		& $\sim 4.29\times 10^{12}$ & \\
10		& $\sim 8.20\times 10^{11}$ & \\
9		& 105,984,823,653 	& 2,208,511,226 \\
8		& 10,804,681,959 	& 225,242,556 \\
7		& 932,651,938 		& 19,466,575 \\
6		& 70,763,560 		& 1,482,686 \\
5		& 4,807,552 		& 101,983 \\
4		& 294,507 			& 6,538 \\
3		& 16,204 			& 425 \\
2		& 784 				& 33 \\
1		& 32				& 4 \\
0   	& 1         		& 1 \\ \hline
\end{tabular}
\end{table}

Table \ref{tab1} lists the distribution of the number of permutations that can be realized with 
optimal circuits requiring no more than 9 gates. We estimate the number of functions requiring 
$10..17$ gates using random function size distribution, see Table \ref{tab2}, and optimal 
synthesis of all 3-bit reversible functions. We used $CS1$ to run this test, and it took 
10,549 seconds (under 3 hours) to complete using 43.04 GB of RAM. $CS2$ used 2.74 GB RAM 
and took 743.401 seconds (under 13 minutes) to synthesize optimal implementations with up 
to 8 gates.

\subsection{Optimal linear circuits}
\begin{table}[h]
\centering
\caption{Number of 4-bit linear reversible functions requiring 0..10 gates 
in an optimal implementation.}
\label{tablin}
\begin{tabular}{|r|r|} \hline
Size  	& Functions  \\ \hline
10		& 138 \\
9		& 13555 \\
8		& 84225 \\
7		& 118424 \\
6		& 72062 \\
5		& 26182 \\
4		& 6589 \\
3		& 1206 \\
2		& 162 \\
1		& 16 \\
0   	& 1 \\ \hline
\end{tabular}
\end{table}

Linear reversible circuits are the most complex part of error correcting circuits \cite{ar:ag}. Efficiency of these circuits 
defines efficiency 
of quantum encoding and decoding error correction operations.
Linear reversible functions are those whose positive polarity Reed-Muller polynomial has only linear terms. More simply, 
linear reversible functions are those computable by circuits with NOT and CNOT gates. 

For example, the reversible mapping 
$a,b,c,d \mapsto b\oplus 1,a \oplus c\oplus 1,d\oplus 1,a$ is a linear reversible function. Interestingly, this linear function 
is one of the 138 
most complex linear reversible functions---it requires 10 gates in an optimal implementation. The optimal implementation of this 
function 
is given by the circuit CNOT(b,a) CNOT(c,d) CNOT(d,b) NOT(d) CNOT(a,b) CNOT(d,c) CNOT(b,d) CNOT(d,a) NOT(d) CNOT(c,b). 

We synthesized optimal circuits for all 322,560 4-bit linear reversible functions.  This process took under two seconds on 
$CS2$. The distribution 
of the number of functions requiring a given number of gates is shown in Table \ref{tablin}.

\subsection{Synthesis of Benchmarks}

\begin{table}[t]
\centering
\caption{Optimal implementations of benchmark functions.}
\label{tab3}
{\footnotesize 
\begin{tabular}{|r|r||r|r|r||r|r|r|} \hline
Name       & Specification         & SBKC     & Source  & PO?    & SOC    & Our optimal circuit & Runtime \\  \hline

4\_49      & [15,1,12,3,5,6,8,7,    & 12    & \cite{www:rlsb} & No    & 12    & NOT(a) CNOT(c,a) CNOT(a,d) TOF(a,b,d)  &  
.000690s  \\
              & 0,10,13,9,2,4,14,11] &        &                    &     &         & CNOT(d,a) TOF(c,d,b) TOF(a,d,c) TOF(b,c,a)  
&   \\
             &&&&&                                                            & TOF(a,b,d) NOT(a) CNOT(d,b) CNOT(d,c) & \\\hline

4bit-7-8   & [0,1,2,3,4,5,6,8,7,9,  & 7        & \cite{co:m}      & No    & 7        & CNOT(d,b) CNOT(d,a) CNOT(c,d) 
TOF4(a,b,d,c) & .000003s \\
             & 10,11,12,13,14,15] &         &                   &     &         & CNOT(c,d) CNOT(d,b) CNOT(d,a) &   \\\hline

decode42   & [1,2,4,8,0,3,5,6,7,9,& 11    & \cite{ar:gaj}      & No    & 10    & CNOT(c,b) CNOT(d,a) CNOT(c,a) TOF(a,d,b)  &  
.000006s \\
             & 10,11,12,13,14,15] &       &                    &       &       & CNOT(b,c) TOF4(a,b,c,d) TOF(b,d,c)  &   \\
             &&&&&                                                            & CNOT(c,a) CNOT(a,b) NOT(a) & \\\hline

hwb4       & [0,2,4,12,8,5,9,11,1,& 11    & \cite{www:rlsb} & Yes & 11    & CNOT(b,d) CNOT(d,a) CNOT(a,c) TOF4(b,c,d,a) & 
.000106s \\
             & 6,10,13,3,14,7,15]     &       &                   &     &       & CNOT(d,b) CNOT(c,d) TOF(a,c,b) TOF4(b,c,d,a) &   
\\
             &&&&&                                                            & CNOT(d,c) CNOT(a,c) CNOT(b,d) & \\\hline

imark      & [4,5,2,14,0,3,6,10,  & 7        & \cite{ar:pspmh} & No    & 7        & TOF(c,d,a) TOF(a,b,d) CNOT(d,c) CNOT(b,c) & 
.000003s   \\
             & 11,8,15,1,12,13,7,9] &         &                   &      &          & CNOT(d,a) TOF(a,c,b) NOT(c) &   \\\hline

mperk      & [3,11,2,10,0,7,1,6,  & 9*    & \cite{ws:p,co:m}& No    & 9        & NOT(c) CNOT(d,c) TOF(c,d,b) TOF(a,c,d) & 
.000003s \\
             & 15,8,14,9,13,5,12,4] &       &                   &        &         & CNOT(b,a) CNOT(d,a) CNOT(c,a) CNOT(a,b) &   
\\
             &&&&&                                                            & CNOT(b,c) & \\\hline

oc5        & [6,0,12,15,7,1,5,2,4, & 15    & \cite{co:szs}      & No    & 11    & TOF(b,d,c) TOF(c,d,b) TOF(a,b,c) NOT(a) & 
.000313s \\
             & 10,13,3,11,8,14,9] &          &                   &       &       & CNOT(d,b) CNOT(a,c) TOF(b,c,d) CNOT(a,b) & \\
             &&&&&                                                            & CNOT(c,a) CNOT(a,c) TOF4(a,b,d,c) & \\\hline

oc6        & [9,0,2,15,11,6,7,8, & 14        & \cite{co:szs}      & No    & 12    & TOF4(b,c,d,a) TOF4(a,c,d,b) CNOT(d,c) 
TOF(b,c,d) & .000745s \\
             & 14,3,4,13,5,1,12,10] &         &                   &      &         & TOF(c,d,a) TOF4(a,b,d,c) CNOT(b,a) NOT(a) & 
\\
             &&&&&                                                            & CNOT(c,b) CNOT(d,c) CNOT(a,d) TOF(b,d,c) & 
\\\hline

oc7        & [6,15,9,5,13,12,3,7, & 17    & \cite{co:szs}      & No    & 13    & TOF(b,d,c) TOF(a,b,d) CNOT(b,a) TOF4(a,c,d,b) & 
.0265s \\
             & 2,10,1,11,0,14,4,8] &           &                   &     &          & CNOT(c,b) CNOT(d,c) TOF(a,c,d) NOT(b) 
NOT(d) & \\
             &&&&&                                                            & CNOT(b,c) TOF(b,d,a) TOF(a,c,d) CNOT(c,a) & 
\\\hline

oc8        & [11,3,9,2,7,13,15,14, & 16    & \cite{co:szs}      & No    & 12    & CNOT(d,a) TOF(b,c,a) TOF(c,d,b) TOF4(a,b,d,c) 
& .001395s \\
             & 8,1,4,10,0,12,6,5] &          &                   &      &          & TOF(a,b,d) TOF(a,d,b) NOT(a) NOT(b) & \\
             &&&&&                                                            & TOF(b,d,a) CNOT(a,d) TOF(b,c,d) & \\\hline

primes4    & [2,3,5,7,11,13,0,1,4, & N/A    & N/A        & N/A    & 10    & CNOT(d,c) CNOT(c,a) CNOT(b,c) NOT(b) & .000012s \\
             & 6,8,9,10,12,14,15] &            &                  &        &        & TOF(b,c,d) TOF4(a,b,d,c) TOF(a,c,b) & \\
             &&&&&                                                            &  NOT(a) TOF4(a,c,d,b) CNOT(b,a) & \\\hline

rd32       & [0,7,6,9,4,11,10,13, & 4        & \cite{ar:f}      & Yes & 4     & TOF(a,b,d) CNOT(a,b) TOF(b,c,d) CNOT(b,c) &  
.000002s \\
             & 8,15,14,1,12,3,2,5] &          &                   &     &          &  &   \\ \hline

shift4     & [1,2,3,4,5,6,7,8,9, & 4         & \cite{co:m}      & Yes    & 4        & TOF4(a,b,c,d) TOF(a,b,c) CNOT(a,b) NOT(a) 
& .000002s \\
             & 10,11,12,13,14,15,0] &       &                   &     &          &  &   \\ \hline
\end{tabular}
}
\end{table}

In this subsection, we present optimal circuits for benchmark functions that have been previously reported in the literature. 
Table \ref{tab3} summarizes the results.  The table describes the {\bf Name} of the benchmark function, its complete {\bf 
Specification}, {\bf S}ize of the {\bf B}est {\bf K}nown {\bf C}ircuit ({\bf SBKC}), the {\bf Source} of this circuit, indicator 
of whether this circuit has been {\bf P}roved {\bf O}ptimal ({\bf PO?}), {\bf S}ize of an {\bf O}ptimal {\bf C}ircuit ({\bf 
SOC}), the optimal implementation that our program found, and the runtime our program takes to find this optimal implementation.  
We used $CS1$ for this test, and report the runtime it takes after hash table with all optimal implementations with up to 9 
gates is loaded into RAM.  Shorter runtimes were identified using multiple runs of the search to achieve sufficient accuracy.  
Please note that we introduce the function $primes4$, which cannot be found in previous literature. Also, the 9-gate circuit for 
function {\em mperk} requires some extra SWAP gates to properly map inputs into their respective outputs, indicated by an 
asterisk.

\subsection{Searching for a Hard Permutation}

We executed a 12-hour search using $CS1$ to find a permutation requiring more than 14 gates in an optimal implementation.
To run the search, we used 14- and 13- gate optimal implementations and tried to extend them by assigning gates to the beginning 
and
the end of those optimal implementations, computing the resulting function, and verifying how many gates they require. After the 
12 hour search, we were
not able to find a permutation requiring more than 14 gates, indicating further that there are not many such permutations.

%
%

\section{Conclusions and Future Work} \label{s:cfr}
In this paper, we described an algorithm that finds an optimal circuit for any 4-bit reversible function. Our goal was to 
minimize the number of gates required for function implementation. Our program implementation takes approximately 3 hours to 
calculate all optimal implementations requiring up to 9 gates, and then an average of about $0.01$ seconds to search for an 
optimal circuit of any 4-bit reversible function. Both calculations are surprisingly fast given the size of the search space.

We demonstrated the synthesis of 117,798,040,190 optimal circuits in 10,549 seconds, amounting to an average speed of 11,166,749 
circuits per second.  This is over 65 times faster and some 4,500 times more than the best previously reported result (26 
million circuits in 152 seconds)~\cite{ar:pspmh}.
We synthesized optimal implementations for all linear reversible functions.

We also demonstrated that the search for an optimal circuit can be done very quickly. For example, if all optimal circuits are 
written to a {\em hypothetical} 100+TB 5400 RPM hard drive, the average time to extract a random circuit from the drive would be 
expected to take on the order of $0.01-0.02$ seconds (typical access time for 5400 RPM hard drives). In other words, it would 
likely take longer to read the answer from a {\em hypothetical} hard drive than to compute it with our implementation. 
Furthermore, the 3-hour calculation of all optimal circuits with up to 9 gates could be reduced by storing its result (computed 
once for the entirety of the described search and its follow up executions) on the hard drive, as was done in Subsection 
\ref{ss:1}. It took 1111 seconds, i.e., under 18 minutes, to load optimal circuits with up to 9 gates into RAM using $CS1$. 
Given that the media
transfer rate of modern hard drives is 1Gbit/s (=1GB in 8 seconds) and higher, it may take no longer than 5 minutes ($=300$s $> 
296=37*8$s) to load optimal implementations into RAM to initiate the search on a different machine.

Minor modifications to the algorithm could be explored to address other optimization issues.  For example, for practicality, one 
may be interested in minimizing depth.  This may be important if a faster circuit is preferred, and/or if quantum noise has a 
stronger constituent with time, than with the disturbance introduced by multiple gate applications.  It may also be important to 
account for the different implementation costs of the gates used (generally, NOT is much simpler than CNOT, which in turn, is 
simpler than Toffoli). Both modifications are possible, by making changes only to the first part of the search.

To optimize depth, one needs to consider a different family of gates, where, for instance, sequence NOT$(a)\;$CNOT$(b,c)$ is 
counted as a single gate. To account for different gate costs, one needs to search for small circuits via increasing cost by one 
(assuming costs are given as natural numbers), as opposed to adding a gate to all maximal size optimal circuits.

It is also possible to extend the search to find optimal implementations in restricted architectures (trivially if an optimal 
implementation is required up to the input/output permutation). Finally, the search could be extended to find some small optimal 
5-bit circuits. However, more work is necessary to determine how far the search progress could be carried with 5-bit optimal 
implementations.

Our future plans include:
\begin{itemize}
\item execution of an extended search to find a hard permutation, one requiring a large ($\geq 15$) number of gates;
\item construction of a representative set of functions that could be used to test heuristic synthesis algorithms against;
\item computing all numbers in Table \ref{tab1} exactly. If completed, this automatically solves the first
task by appropriately keeping track of relevant computations. Also, this would give an exact number for $L(4)$, the maximal 
number of gates required to implement a 4-bit reversible function, and help with the second task;
\item finding depth-optimal 4-bit circuits and optimal 5-bit circuits. In particular, a simple calculation shows that using 
$CS1$
it is possible to compute all optimal 5-bit circuits with up to six gates, and thus it is possible to search optimal 5-bit 
implementations with up to 12 gates;
\item extending techniques reported in this paper to the synthesis of optimal stabilizer circuits. Coupled with peep-hole 
optimization algorithm for circuit simplification, these results may become a very useful tool in optimizing error correction 
circuits. This may be of a particular practical interest since implementations of quantum algorithms may be expected to be 
dominated by error correction circuits. 
\end{itemize}

\section{Acknowledgments}
We wish to thank Dr. Donny Cheung from the University of Calgary for his useful discussions.
We wish to thank Dr. Barry Schneider from the National Science Foundation for giving us access to his computer, named $CS1$
in our paper.

This article was based on work partially supported by the National Science Foundation, during
D. Maslov's assignment at the Foundation. Any opinion, finding and conclusions or recommendations expressed
in this material are those of the author and do not necessarily reflect the views of the National
Science Foundation.
%

\bibliographystyle{abbrv}

\end{document}